\documentclass{article}
\usepackage{nips2004e,times}
\usepackage{epsf}
\usepackage{graphicx}
\usepackage{psfig}
\usepackage{epsfig}
\usepackage{rotating}
\usepackage{amssymb}
\usepackage{amsmath}
\usepackage{subfigure}
\title{Comparing Beliefs, Surveys and Random Walks}
\author{
Erik Aurell\\
SICS, Swedish Institute of Computer Science\\
P.O. Box 1263, SE-164 29 Kista, Sweden \\
and Dept. of Physics,  \\
KTH -- Royal Institute of Technology \\
AlbaNova -- SCFAB SE-106 91 Stockholm, Sweden\\
\texttt{eaurell@sics.se} \\
\And
Uri Gordon and Scott Kirkpatrick \\
School of Engineering and Computer Science \\
Hebrew University of Jerusalem \\
91904 Jerusalem, Israel \\
\texttt{\{guri,kirk\}@cs.huji.ac.il} \\
}

\begin{document}

\maketitle

\begin{abstract}
Survey propagation is a powerful technique from statistical physics that has been
applied to solve the 3-SAT problem both in principle and in practice. We give, using
only probability arguments, a common derivation of survey propagation, belief
propagation and several interesting hybrid methods. We then present numerical
experiments which use WSAT (a widely used random-walk based SAT solver) to quantify
the complexity of the 3-SAT formulae as a function of their parameters, both as
randomly generated and after simplification, guided by survey propagation.  Some
properties of WSAT which have not previously been reported make it an ideal tool for
this purpose -- its mean cost is proportional to the number of variables in the
formula (at a fixed ratio of clauses to variables) in the easy-SAT regime and
slightly beyond, and its behavior in the hard-SAT regime appears to reflect the
underlying structure of the solution space that has been predicted by replica
symmetry-breaking arguments.  An analysis of the tradeoffs between the various
methods of search for satisfying assignments shows WSAT to be far more powerful than
has been appreciated, and suggests some interesting new directions for practical
algorithm development.
\end{abstract}
\section{Introduction}
Random 3-SAT is a classic problem in combinatorics, at the heart of computational
complexity studies and a favorite testing ground for both exactly analyzable 
and heuristic solution methods which are then applied to a wide variety of problems 
in machine learning and artificial intelligence.  It consists of a ensemble of 
randomly generated logical expressions, each depending on $N$ Boolean variables 
$x_i$, and constructed by taking the AND of $M$ clauses.  Each clause $a$ consists of the 
OR of 3 ``literals'' $y_{i,a}$. $y_{i,a}$ is taken to be either $x_i$ or $\neg x_i$ 
at random with equal probability, and the three values of the index $i$ in each 
clause are distinct.
Conversely, the neighborhood of a variable $x_i$ is $V_i$, the set of
all clauses in which $x_i$ or $\neg x_i$ appear.  
For each such random formula, one asks whether there is some 
set of $x_i$ values for which the formula evaluates to be TRUE. The ratio 
$\alpha = M/N$ controls the difficulty of this decision problem, and predicts 
the answer with high accuracy, at least as both $N$ and $M$ tend to infinity, 
with their ratio held constant.  At small $\alpha$, solutions are easily found, 
while for sufficiently large $\alpha$ there are almost certainly no satisfying 
configurations of the ${x_i}$, and compact proofs of this fact can be constructed.
Between these limits lies a complex, 
spin-glass-like phase transition, at which the cost of analyzing the problem with 
either exact or heuristic methods explodes.  

A recent series of papers drawing upon the statistical mechanics of disordered 
materials has not only clarified the nature of this transition, but also lead to 
a thousand-fold increase in the size of the concrete problems that can be 
solved~\cite{MPZ,MZ,Braunstein-Mezard-Zecchina}
This paper provides a derivation of the new methods using 
nothing more complex than probabilities, suggests some generalizations, and reports 
numerical experiments that disentangle the contributions of the several component 
heuristics employed. For two related discussions, see ~\cite{ParisiProb, BraunsteinBP}.

An iterative "belief propagation"~\cite{Pearl88} (BP) algorithm for K-SAT can be derived to evaluate the probability, or
"belief," that a variable will take the value TRUE in variable
configurations that satisfy the formula considered.
To calculate this, we first define a message ("transport") sent from a variable to a clause: 
\begin{itemize}
\item $t_{i\to a}$ is 
{\it the probability
that variable $x_i$ satisfies clause $a$}
\end{itemize}
In the other direction, we define a message ("influence") sent from a clause
to a variable:
\begin{itemize}
        \item   $i_{a\to i}$ is
{\it the probability that clause $a$ is satisfied by another variable than $x_i$}
\end{itemize}
In 3-SAT, where clause $a$ depends on variables $x_i$, $x_j$ and $x_k$,
BP gives the following iterative update equation for its influence.
\begin{eqnarray}
\label{eq:Scott-scaled-i-iterations}
i^{(l)}_{a\to i} &=& t^{(l)}_{j\to a} + t^{(l)}_{k\to a} -
                t^{(l)}_{j\to a} t^{(l)}_{k\to a}
\end{eqnarray}

The BP update equations for the transport $t_{i\to a}$
involve the products of influences acting on a variable
from the clauses which surround $x_i$, forming its "cavity," $V_i$, sorted by which literal ($x_i$ or $\neg x_i$)
appears in the clause:
\begin{equation}
\label{eq:A}
A^0_i = \prod_{b\in V_i,\, y_{i,b} = \neg x_i} i_{b\to i} \qquad \hbox{and} \qquad
A^1_i = \prod_{b\in V_i,\, y_{i,b} = x_i} i_{b\to i}
\end{equation}
The update equations are then
\begin{equation}
\label{eq:Scott-scaled-t-iterations}
t^{(l)}_{i\to a} = \left\{ \begin{array}{ll}
\frac{i^{(l-1)}_{a\to i}A^1_i} {i^{(l-1)}_{a\to i}A^1_i + A^0_i } \quad &\hbox{if
$y_{i,a} = \neg x_i$}\\
&\quad\\
\frac{i^{(l-1)}_{a\to i}A^0_i} {i^{(l-1)}_{a\to i}A^0_i + A^1_i }    
&\hbox{if $y_{i,a} = x_i$ }
\end{array}\right.
\end{equation}
The superscripts $(l)$ and $(l-1)$ denote iteration.
The probabilistic interpretation is the following:
suppose we have $i^{(l)}_{b\to i}$ for all clauses $b$
connected to variable $i$. Each of these clauses can
either be satisfied by another
variable (with probability $i^{(l)}_{b\to i}$), or not
be satisfied by another variable (with probability $\left(1-i^{(l)}_{b\to i}\right)$),
and also be satisfied by variable $i$ itself.
If we set variable $x_i$ to 0, then some clauses are
satisfied by $x_i$, and some have to be satisfied by other variables.
The probability that they are all
{\it satisfied} is $\prod_{b\neq a, y_{i,b} = x_i} i^{(l)}_{b\to i}$.
Similarly, if $x_i$ is set to 1 then all these clauses $b$
are satisfied with
probability $\prod_{b\neq a, y_{i,b}=\neg x_i} i^{(l)}_{b\to i}$.
The products in  (\ref{eq:Scott-scaled-t-iterations}) can therefore be
interpreted as joint probabilities of independent events. 
Variable $x_i$ can be $0$ or $1$ in a solution if the clauses in which 
$x_i$ appears are either satisfied directly by $x_i$ itself, or by other
variables. Hence
\begin{equation}
\label{eq:single-bit-probabilities}
\hbox{Prob}(x_i)= \frac{A^0_i}{A^0_i+A^1_i} \qquad \hbox{and} \qquad
\hbox{Prob}(\neg x_i) = \frac{A^1_i}{A^0_i+A^1_i}
\end{equation}
A BP-based decimation scheme results from fixing the
variables with largest probability to be either true or false.  We then recalculate 
the beliefs for the reduced formula, and repeat.

To arrive at SP we introduce a modified system of beliefs:  every
variable falls into one of three classes: TRUE in all solutions (1);
FALSE in all solutions(0); and
TRUE in some and FALSE in other solutions ($free$).
The message from a clause to a variable (an influence) is then
the same as in BP above.
Although we will again only need to keep track of one message
from a variable to a clause (a transport), it is convenient
to first introduce three ancillary messages: 
\begin{itemize}
        \item $\hat T_{i\to a}(1)$ is {\it the probability
that variable $x_i$ is true in clause $a$ in all solutions}
        \item $\hat T_{i\to a}(0)$ is {\it the probability
that variable $x_i$ is false in clause $a$ in all solutions}
        \item $\hat T_{i\to a}(free)$ is {\it the probability
that variable $x_i$ is true in clause $a$ in some solutions and false in others}.
\end{itemize}
Note that there are here three transports 
for each directed link
$i\to a$, from a variable to a clause, in the graph.
As in BP, these numbers
will be functions of the influences from clauses to variables
in the preceeding update step.
Taking again the incoming influences independent, we have
\begin{equation}
\label{eq:three-states}
\begin{array}{lclr}
\hat T^{(l)}_{i\to a}(free) &\propto&
        \prod_{b\in V_i\setminus a} &i^{(l-1)}_{b\to i}\\
\hat T^{(l)}_{i\to a}(0)  +
\hat T^{(l)}_{i\to a}(free)&\propto&\prod_{b\in V_i\setminus a, y_{i,b} =  x_i}
&i^{(l-1)}_{b\to i} \\
\hat T^{(l)}_{i\to a}(1) +
\hat T^{(l)}_{i\to a}(free) &\propto&
        \prod_{b\in V_i\setminus a, y_{i,b} = \neg x_i} &i^{(l-1)}_{b\to i}
\end{array}
\end{equation}
The proportionality indicates that the probabilities are
to be normalized. We see that the structure is quite similar
to that in BP. 
But we can make it closer still by introducing
$t_{i\to a}$ with the same meaning as in BP.
In SP it will then, as the case might be, be equal to
to $T_{i\to a}(free) +T_{i\to a}(0)$
or $T_{i\to a}(free) +T_{i\to a}(1)$.
That gives (compare (\ref{eq:Scott-scaled-t-iterations})):
\begin{equation}
\label{eq:SP-final-t-iterations}
t^{(l)}_{i\to a} = \left\{ \begin{array}{ll}
\frac{i^{(l-1)}_{a\to i}A^1_i} {i^{(l-1)}_{a\to i}A^1_i + A^0_i - A^1_i A^0_i} \quad
&\hbox{if $y_{i,a} = \neg x_i$}\\
&\quad\\
\frac{i^{(l-1)}_{a\to i}A^0_i} {i^{(l-1)}_{a\to i}A^0_i + A^1_i - A^1_i A^0_i } 
&\hbox{if $y_{i,a} = x_i$}
\end{array}\right.
\end{equation}
The update equations for
$t_{i\to a}$ are the same in SP as in BP,
{\i.e.} one uses (\ref{eq:Scott-scaled-i-iterations}) in SP as well.  
Similarly to (\ref{eq:single-bit-probabilities}), decimation
now removes the most fixed variable, i.e. the one with the
largest absolute value of
$(A^0_i-A^1_i)/(A^0_i + A^1_i - A^1_i A^0_i )$.
Given the complexity of the original
derivation of SP~\cite{MPZ,MZ}, it is remarkable that the SP
scheme can be interpreted as a type of belief propagation in
another belief system. And even more remarkable  that the
final iteration formulae differ so little.

A modification of SP which we will consider in the
following is to interpolate between BP $(\rho=0)$ and SP
$(\rho=1)$~\footnote{This interpolation has also been considered and implemented by 
R. Zecchina and co-workers.}
by considering equations
\begin{equation}
\label{eq:SP-BP-interpolation-iterations}
t^{(l)}_{i\to a} \left\{ \begin{array}{ll}
\frac{i^{(l-1)}_{a\to i}A^1_i} {i^{(l-1)}_{a\to i}A^1_i + A^0_i - \rho A^1_i A^0_i}
\quad &\hbox{if $y_{i,a} = \neg x_i$}\\
&\quad\\
\frac{i^{(l-1)}_{a\to i}A^0_i} {i^{(l-1)}_{a\to i}A^0_i + A^1_i -\rho A^1_i A^0_i }\
&\hbox{if $y_{i,a} = x_i$}
\end{array}\right.
\end{equation}
We do not have an interpretation of the intermediate cases of $\rho$ as belief systems.

\section{The Phase Diagram of 3-SAT} 

Early work on developing 3-SAT heuristics discovered that as $\alpha$ is 
increased, the problem changes from being easy to solve to extremely hard, then 
again relatively easy when the formulae are almost certainly UNSAT.  It was natural 
to expect that a sharp phase boundary between SAT and UNSAT phases in the limit of 
large $N$ accompanies this ``easy-hard-easy'' observed transition, and the finite-size 
scaling results of \cite{KirkpatrickSelman94} confirmed this. Their work placed the 
transition at about $\alpha = 4.2$.  Monasson and Zecchina
\cite{MonassonZecchina} soon showed, using the replica method from statistical 
mechanics, that the phase transition to be expected had unusual characteristics,
including 
``frozen variables'' and a highly nonuniform distribution of solutions, making 
search difficult.  Recent technical advances have made it possible to use simpler 
cavity mean field methods to pinpoint the SAT/UNSAT boundary at $\alpha = 4.267$ and 
suggest that the ``hard-SAT'' region in which the solution space becomes inhomogeneous 
begins at about $\alpha = 3.92$. 
These calculations also predicted a specific solution structure 
(termed 1-RSB for ``one step replica symmetry-breaking'') \cite{MPZ,MZ}
in which the 
satisfiable configurations occur in large clusters, maximally separated 
from each other.  Two types of frozen variables are predicted, one set 
which take the same value in all clusters and a second set whose value 
is fixed within a particular cluster.  The remaining variables are
``paramagnetic'' and can take either value in some of the states of a given 
cluster.  A careful analysis of the 1-RSB solution has subsequently shown that this 
extreme structure is only stable above $\alpha = 4.15$.  Between 3.92 and 4.15 a wider range of 
cluster sizes, and wide range of inter-cluster Hamming 
distances are expected\cite{MontanariParisiRicci-Tersinghi03}. As a result, 
we expect the values $\alpha =$ $3.9$, $4.15$ and $4.267$ to separate regions in which 
the nature of the 3-SAT decision problem is distinctly different.

\begin{center}
\begin{figure}
\label{fig:dependance-on-rho.}
\epsfig{figure=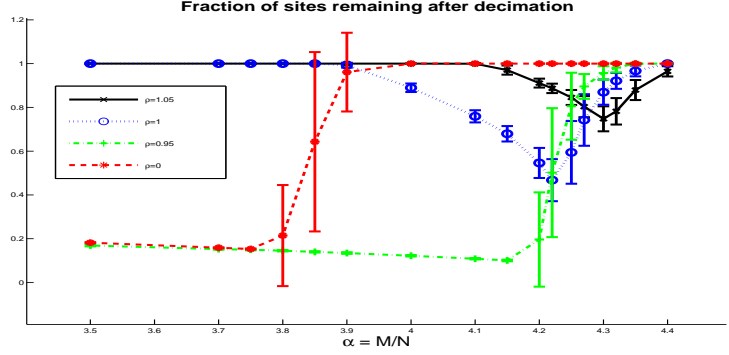,height=12truecm,width=5truecm,clip=, angle=-90}
\caption{Dependence of decimation depth on the interpolation parameter $\rho$.}
\end{figure}
\end{center}
``Survey-induced decimation'' consists of using 
SP to determine the variable most likely to be frozen, then setting that variable 
to the indicated frozen value, simplifying the formula as a result, updating the 
SP calculation, and repeating the process.  For $\alpha < 3.9$ we 
expect SP to discover 
that all spins are free to take on more than one value in some 
ground state, so no spins will be decimated.  Above $3.9$, SP ideally should 
identify frozen spins until all that remain are paramagnetic.  The depth of 
decimation, or fraction of spins reminaing when SP sees only paramagnetic spins, 
is thus an important characteristic.  We show in Fig. 1 the fraction of spins 
remaining after survey-induced decimation for values of $\alpha$ from $3.85$ to $4.35$ in 
hundreds of formulae with $N = 10,000$.  The error bars show the standard 
deviation, which becomes quite large for large values of $\alpha$.  To the left of 
$\alpha = 4.2$, on the descending part of the curves, SP reaches a paramagnetic 
state and halts.  On the right, or ascending portion of the curves, SP stops by 
simply failing to converge.

Fig 1 also shows how different the behavior of BP and the hybrids between BP and SP 
are in their decimation behavior.  We studied BP ($\rho = 0$), 
underrelaxed SP ($\rho = 0.95$), SP, and overrelaxed SP 
($\rho = 1.05$). BP and underrelaxed SP do not reach a paramagnetic state, 
but continue until the formula breaks apart into clauses that have no variables 
shared between them.  We see in Fig.~1 that BP stops working at roughly 
$\alpha = 3.9$, the point at which SP begins to operate.  The underrelaxed SP 
behaves like BP, but can be used well into the RSB region.  
On the rising parts 
of all four curves in Fig 1, the scheme halted as the surveys ceased 
to converge.  Overrelaxed SP in Fig.~1
may give reasonable recommendations 
for simplification even on formulae which are likely to be UNSAT. 
   
\section{Some Background on WSAT}

Next we consider WSAT, the random walk-based search routine used to finish the job
of exhibiting a satisfying configuration after SP (or some other decimation advisor)
has simplified the formula.  The surprising power exhibited by SP has to some extent
obscured the fact that WSAT is itself a very powerful tool for solving constraint
satisfaction problems, and has been widely used for this.  Its running time,
expressed in the number of walk steps required for a successful search is also useful
as an informal definition of the complexity of a logical formula.  Its history goes
back to Papadimitriou's~\cite{Papadimitriou91}
observation that
a subtly biased random walk would with high probability discover satisfying solutions in the simpler 2-SAT problem after, at worst, $O(N^2)$ steps.  His procedure was to start
with an arbitary assignment of values to the binary variables, then reverse the sign
of one variable at a time using the following random process:
\begin{itemize}  
\item        select an unsatisfied clause at random
\item        select at random a variable that appears in the clause
\item        reverse that variable
\end{itemize}

This procedure, sometimes called RWalkSAT, works because changing the sign of a
variable in an unsatisfied clause always satisfies that clause and, at first, has no
net effect on other clauses.  It is much more powerful than was proven
initially.  Two recent papers~\cite{SemerjianMonasson03,BarthelHartmannWeigt03}.
have argued analytically and shown experimentally that Rwalksat finds satisfying
configurations of the variables after a number of steps that is proportional to $N$
for values of $\alpha$ up to roughly $2.7$, after which this cost increases
exponentially with $N$.  

\begin{figure}[htb!]
\begin{center}
\label{fig:2}
\begin{tabular}{ccc}
\epsfig{figure=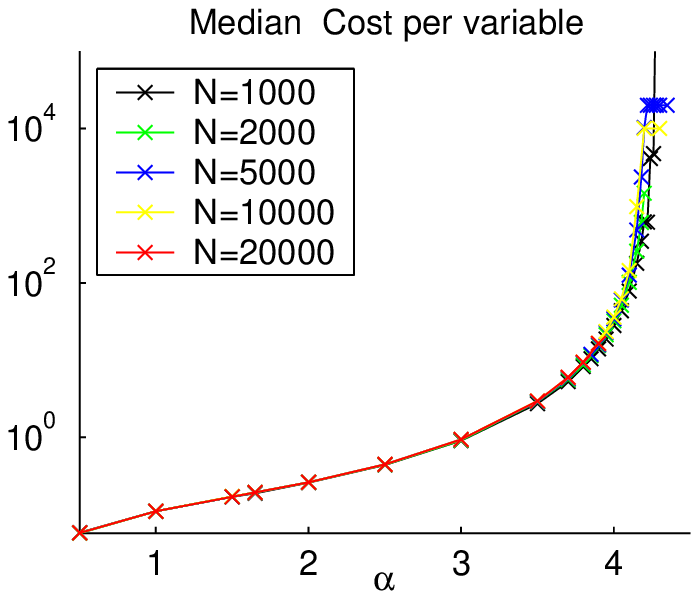,,height=4truecm,width=5truecm,clip=, angle=0} &
\epsfig{figure=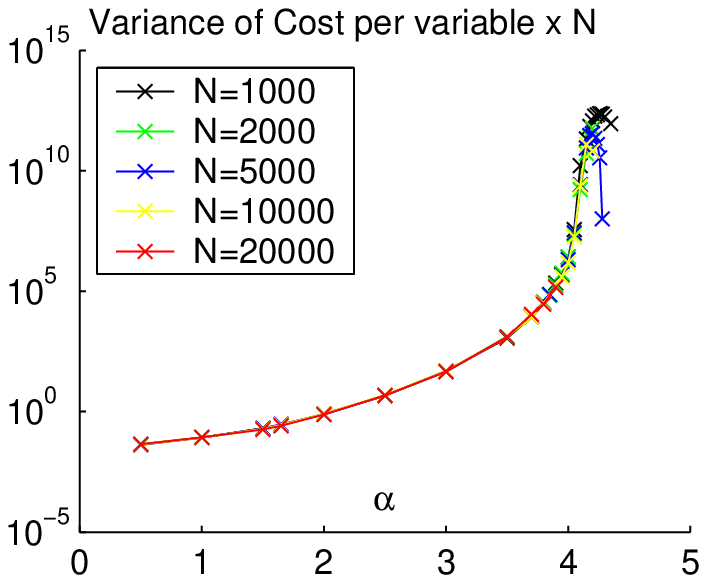,,height=4truecm,width=5truecm,clip=, angle=0} &
\hspace{0.5cm}
\end{tabular}
\caption{(a) Median of WSAT cost per variable in 3-SAT as a function of $\alpha$.
(b) Variance of WSAT cost, scaled by $N$.}
\end{center}
\end{figure}

The second trick in WSAT was introduced by Kautz and Selman~\cite{SKwalksat}.  They also
choose an unsatisfied clause at random, but then reverse one of the ``best'' variables, selected
at random, where ``best'' is defined as causing the fewest satisfied
clauses to become unsatisfied.  For robustness, they mix this greedy move with
random moves as used in RWalkSAT, recommending an equal mixture of the two types of moves.  Barthel {\it et al.}\cite{BarthelHartmannWeigt03}
used these two moves in 
numerical experiments, but found little improvement over RWalkSAT.

There is a third trick in the most often used variant of WSAT, introduced slightly
later~\cite{SKCwalksat}.  If any variable in the selected unsatisfied clause can be reversed without causing any
other clauses to become unsatisfied, this ``free'' move is immediately accepted and no
further exploration is required. Since we shall show that WSAT works well above
$\alpha = 2.7$, this third move apparently gives WSAT its extra power. 
Although these moves were chosen by the authors of WSAT after considerable
experiment, we have no insight into why they should be the best choices.  

In Fig. 2a, we show the median number of random walk steps per variable taken by the
standard version of WSAT to solve 3-SAT formulas at values of $\alpha$ ranging from
0.5 to 4.3 and for formulae of sizes ranging from $N = 1000$ to $N = 20000$.  The
cost of WSAT remains linear in $N$ well above $\alpha = 3.9$. WSAT cost
distributions were collected on at least 1000 cases at each point.  Since the
distributions are asymmetric, with strong tails extending to higher cost, it is not
obvious that WSAT cost is, in the statistical mechanics language, self-averaging, or
concentrated about a well-defined mean value which dominates the distribution as $N
\rightarrow \infty$.  
To test this, we calculated higher moments of the WSAT cost
distribution and found that they scale with simple powers of N.  For example, in
Fig. 2b, we show that the variance of the WSAT cost per variable, scaled up by N, is a well-defined
function of $\alpha$ up to almost 4.2.  The third and fourth moments of the
distribution (not shown) also are constant when multiplied by $N$ and by $N^2$, respectively.  The WSAT
cost per variable is thus given by a distribution which concentrates with increasing N in
exactly the way that a process governed by the usual laws of large numbers is
expected to behave, even though the typical cost increases by six orders of
magnitude as we move from the trivial cases to the critical regime.     

A detailed analysis of the cost distributions which we observed will be published
elsewhere but we conclude that the median cost of solving 3-SAT using the WSAT
random walk search, as well as the mean cost if that is well-defined, remains linear
in $N$ up to $\alpha = 4.15$, coincidentally the onset of 1-RSB. In the 1-RSB
regime, the WSAT cost per variable distributions shift to higher values as $N$
increases, and an exponential increase in cost with $N$ is likely. 
Is 4.15 really the endpoint for WSAT's linearity, or will the search 
cost per variable converge at still larger values of $N$ which we could 
not study?  We define a rough estimate of $N_{onset}(\alpha) $ by study 
of the cumulative distributions of WSAT cost as the value of N for a 
given $\alpha$ above which the distributions cross at a fixed percentile.  
Plotting $\log(N_{onset})$ against $\log (4.15 - \alpha)$ in Fig. 3, we find 
strong indication that 4.15 is indeed an asymptote for WSAT.

\begin{figure}[htb!]
\begin{center}
\label{fig:3}
\begin{tabular}{ccc}
\epsfig{figure=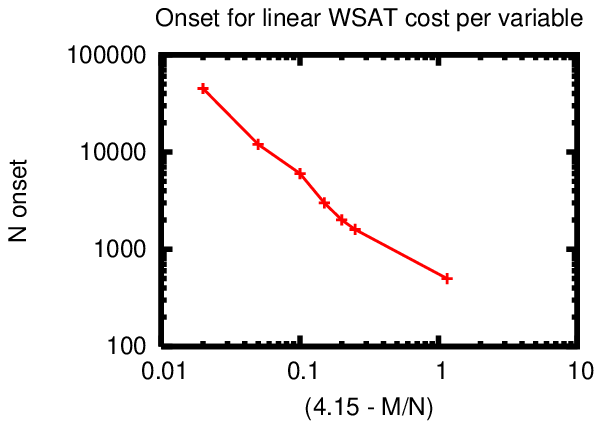,,height=4truecm,width=5truecm,clip=, angle=0} &
\epsfig{figure=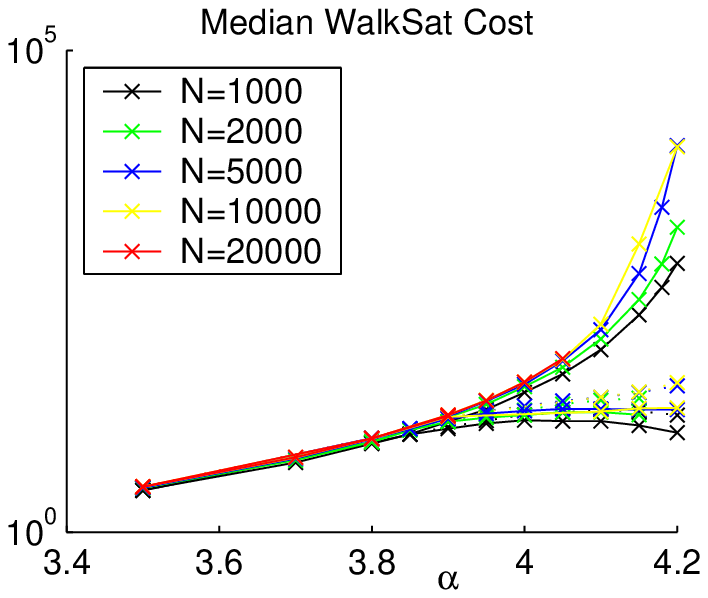,,height=4truecm,width=5truecm,clip=, angle=0} &
\hspace{0.5cm}
\end{tabular}
\caption{Size $N$ at which WSAT cost is linear in N as function of  $4.15 - \alpha$.\\
Figure 4:  WSAT cost, before and after SP-guided decimation.}
\end{center}
\end{figure}

\section{Practical Aspects of SP + WSAT}

The power of SP comes from its use to guide decimation by identifying spins which
can be frozen while minimally reducing the number of solutions that can be
constructed.  To assess the complexity of the reduced formulae that decimation
guided in this way produces we compare, in Fig. 4, the median number of WSAT
steps required to find a satisfying configuration of the variables before and after
decimation.  To a rough approximation, we can say that SP caps the cost of finding a
solution to what it would be at the entry to the critical regime.  There are two
factors, the reduction in the number of variables that have to be searched, and the
reduction of the distance the random walk must traverse when it is restricted to a
single cluster of solutions.  In Fig. 2c the solid lines show the WSAT costs divided by  N, the original number of variables in each formula.  If we instead divide the WSAT cost
after decimation by the number of variables remaining, the complexity measure that
we obtain is only a factor of two larger, as shown by the dotted lines. The relative cost of running WSAT without benefit of decimation is 3-4 decades larger. 

We measured the actual compute time consumed in survey propagation
and in WSAT.  For this we used the Zecchina group's version 1.3 survey propagation
code, and the copy of WSAT (H. Kautz's release 35, see~\cite{walksat-WWW}) 
that they have also employed. 
All programs were run on a Pentium IV Xeon 3GHz dual processor server with 4GB of
memory, and only one processor busy.  We compare timings from runs on the same 100
formulas with $N = 10000$ and $\alpha = 4.1$ and 4.2 (the formulas are simply
extended slightly for the second case).  In the first case, the 100 formulas were
solved using WSAT alone in 921 seconds. Using SP to guide decimation one variable at
a time, with the survey updates performed locally around each modified variable, the
same 100 formulas required 6218 seconds to solve, of which only 31 sec was spent in
WSAT.

When we increase alpha to 4.2, the situation is reversed.  Running WSAT on 100
formulas with $N = 10000$ required 27771 seconds on the same servers, and would have
taken even longer if about half of the runs had not been stopped by a cutoff without
producing a satisfying configuration.  In contrast, the same 100 formulas were
solved by SP followed with WSAT in 10,420 sec, of which only 300 seconds were spent
in WSAT.  The cost of SP does not scale linearly with $N$, but appears to scale as
$N^2$ in this regime.  We solved 100 formulas with $N = 20,000$ using SP followed by
WSAT in 39643 seconds, of which 608 sec was spent in WSAT.  The cost of running SP
to decimate roughly half the spins has quadrupled, while the cost of the final WSAT
runs remained proportional to $N$.

Decimation must stop short of the paramagnetic state at the highest values of $\alpha$, to avoid having SP fail to converge.  In those cases we found that WSAT could sometimes find satisfying configurations if started slightly before this point.  We also explored partial decimation as a means of reducing the cost of WSAT just below the 1-RSB regime, but found that decimation of small fractions of the variables caused the WSAT running times to be highly unpredictable, in many cases increasing strongly.
As a result, partial decimation does not seem to be a useful approach.

\section{Conclusions and future work}

The SP and related algorithms are quite new, so programming improvements may modify the practical conclusions of the previous section.  However, a more immediate target for future work could be the WSAT algorithms.  Further directing its random choices to incorporate the insights gained from BP and SP might make it an effective algorithm even closer to the SAT/UNSAT transition.

\subsubsection*{Acknowledgments}

We have enjoyed discussions of this work with members of the 
replica and cavity theory community, especially 
Riccardo Zecchina, Alfredo Braunstein, Marc Mezard, Remi Monasson and Andrea Montanari.  This work was performed in the framework of EU/FP6 Integrated Project EVERGROW (www.evergrow.org), and in part during a Thematic Institute supported by the EXYSTENCE EU/FP5 network of excellence.  
E.A. acknowledges support from the Swedish Science Council.
S.K. and U.G. are partially supported by a US-Israeli Binational Science Foundation grant.

\end{document}